\documentclass[letter]{article}

\usepackage{graphicx} 
\usepackage{amsmath} 
\usepackage{amssymb}  

\newcommand{\samp}[1]{{\mathcal S}_{#1}}
\newcommand{\hold}[1]{{\mathcal H}_{#1}}
\newcommand{\E}{{\mathcal E}}
\newcommand{\K}{{\mathcal K}}

\newcommand{\dd}{{\mathrm{d}}}
\newcommand{\ee}{{\mathrm{e}}}
\newcommand{\jj}{{\mathrm{j}}}
\newcommand{\bt}{{\mathrm{bt}}}
\newcommand{\one}{{\mathbf{1}}}
\newcommand{\abcd}[4]{\left[\begin{array}{c|c}#1&#2\\\hline#3&#4\end{array}\right]}
\def\ctod{\mathop{\bf c2d}\nolimits}
\def\lift{\mathop{\bf lift}\nolimits}

\def\blkdiag{\mathop{\bf blkdiag}\nolimits}
\newcommand{\mr}{{\mathrm{mr}}}
\newcommand{\sd}{{\mathrm{sd}}}

\newcommand{\dlift}[1]{{\mathbf L}_{#1}}
\newcommand{\idlift}[1]{{\mathbf L}_{#1}^{-1}}
\newcommand{\us}[1]{\uparrow\!#1}
\newcommand{\ds}[1]{\downarrow\!#1}

\newtheorem{problem}{Problem}
\newtheorem{theorem}{Theorem}
\newtheorem{remark}{Remark}

\title{\LARGE \bf
Optimal Discretization of Analog Filters via Sampled-Data $H^\infty$ Control Theory
}

\author{Masaaki Nagahara and Yutaka Yamamoto%
\thanks{M. Nagahara and Y. Yamamoto are with Graduate School of Informatics, 
        Kyoto University, Sakyo-ku Yoshida-Honmachi, Kyoto 606-8501, JAPAN
        {\tt nagahara@ieee.org}, {\tt yy@i.kyoto-u.ac.jp}}%
}
\date{}

\begin{document}

\maketitle
\thispagestyle{empty}
\pagestyle{empty}

\begin{abstract}
In this article, we propose optimal discretization of analog filters (or controllers)
based on
the theory of sampled-data $H^\infty$ control.
We formulate the discretization problem as minimization of the $H^\infty$ norm of the error system
between a (delayed) target analog filter and a digital system including
an ideal sampler, a zero-order hold, and a digital filter.
The problem is reduced to discrete-time $H^\infty$ optimization via the fast sample/hold
approximation method.
We also extend the proposed method to multirate systems.
Feedback controller discretization by the proposed method is discussed
with respect to stability.
Numerical examples show the effectiveness of the proposed method.
\end{abstract}

\section{INTRODUCTION}
Discretization of analog systems is a fundamental transformation in control and signal processing.
Often an analog (or continuous-time) controller is designed based on standard methods
such as PID control \cite{AstHag01}, and then it is implemented in digital devices after discretization.
In signal processing, an analog filter is discretized to implement it on DSP (digital signal processor),
an example of which is active noise control \cite{EllNel93} where the analog model of the secondary path
is discretized and used for an adaptive filter algorithm realized on DSP.

For discretization of analog filters, 
\emph{step-invariant} transformation \cite[Chap.~3]{CheFra} is conventionally and widely used.
The term ``step-invariant'' comes from the fact that
the discrete-time step response of the discretized filter
is exactly the same as the sampled step response
of the original filter.
By this fact, step-invariant transformation is effective for
sufficiently low-frequency signals.
Another well-known discretization method is \emph{bilinear transformation},
also known as \emph{Tustin's method},
which is based on the trapezoidal rule for approximating the definite integral.
The following are advantages of bilinear transformation:
\begin{itemize}
\item stability and minimum-phase property are preserved,
\item there is no error at DC (direct current) between the frequency responses
of the original filter and the discretized one.
\end{itemize}
If a zero error is preferred at another frequency, 
one can use a \emph{prewarping} technique
for bilinear transformation;
see \cite[Sec.~3.5]{CheFra} for details.

Although these methods are widely used,
there may lead considerable discretization errors for
unexpected signals such as signals that contains high-frequency components.
To solve this, we apply the {\em sampled-data $H^\infty$-optimal
filter design} \cite{Nag10,NagOguYam11,YamNagKha12,YamNagKha12-2}
to the discretization problem.
The proposed design procedure is summarized as follows:
\begin{enumerate}
\item give a signal generator model $F(s)$ as an analog filter for $L^2$ input signals,
\item set a digital system $\K$ that contains a sampler, a hold, and a digital filter (see Fig.~\ref{fig:digital-system}),
\item construct an error system $\E$ between $\K$ and a (delayed) target analog filter $G(s)$
with signal model $F(s)$ (see Fig.~\ref{fig:error-system}),
\item find a digital filter in $\K$ that minimizes the $L^2$-induced norm (or $H^\infty$ norm)
of the error system $\E$.
\end{enumerate}
Since the error system is composed of both analog and digital systems,
the optimization is an infinite dimensional one. To reduce this to a finite dimensional optimization,
we introduce the fast sample/hold approximation method \cite{KelAnd92,YamMadAnd99}.
By this method, the optimal digital filter can be effectively 
obtained by numerical computations.

The remainder of this article is organized as follows.
In Section \ref{sec:problem}, we formulate our discretization problem
as a sampled-data $H^\infty$ optimization.
In Section \ref{sec:conventional}, we review two conventional methods:
step-invariant transformation and bilinear transformation.
In Section \ref{sec:H-inf}, we give a design formula to compute
$H^\infty$-optimal filters.
In Section \ref{sec:multirate}, we extend the design method to
multirate systems.
In Section \ref{sec:feedback}, we discuss controller discretization.
Section \ref{sec:examples} presents design examples to illustrate the effectiveness
of the proposed method.
In Section \ref{sec:conclusions}, we offer concluding remarks.

\subsection*{Notation}
Throughout this article, we use the following notation.
We denote by $L^2[0,\infty)$ the Lebesgue space consisting of all square integrable real functions on $[0,\infty)$.
$L^2[0,\infty)$ is sometimes abbreviated to $L^2$.
The $L^2$ norm is denoted by $\|\cdot\|_2$.
The symbol $t$ denotes the argument of time,
$s$ the argument of Laplace transform,
and $z$ the argument of $Z$ transform.
These symbols are used to indicate whether a signal or a system is of continuous-time or discrete-time;
for example, $y(t)$ is a continuous-time signal,
$F(s)$ is a continuous-time system,
$K(z)$ is a discrete-time system.
The operator $\ee^{-ls}$ with nonnegative integer $l$ denotes continuous-time delay (or shift) operator:
$(\ee^{-ls}y)(t)=y(t-l)$.
$\samp{h}$ and $\hold{h}$ denote the ideal sampler and the zero-order hold
respectively with sampling period $h>0$.
A transfer function with state-space matrices $A,B,C,D$ is denoted by
\[
 \abcd{A}{B}{C}{D} 
  := \begin{cases}
    C(sI-A)^{-1}B+D,&\text{~(continuous-time)}\\
    C(zI-A)^{-1}B+D,&\text{~(discrete-time)}
  \end{cases}
\]
We denote the imaginary number $\sqrt{-1}$ by $\jj$.

\section{PROBLEM FORMULATION}
\label{sec:problem}
In this section, we formulate the problem of optimal discretization.

\begin{figure}[tb]
 \centering
\unitlength 0.1in
\begin{picture}( 28.0000,  4.0000)(  4.0000, -6.0000)
%
\special{pn 8}%
\special{pa 400 400}%
\special{pa 800 400}%
\special{fp}%
\special{sh 1}%
\special{pa 800 400}%
\special{pa 734 380}%
\special{pa 748 400}%
\special{pa 734 420}%
\special{pa 800 400}%
\special{fp}%
%
\special{pn 8}%
\special{pa 800 200}%
\special{pa 1200 200}%
\special{pa 1200 600}%
\special{pa 800 600}%
\special{pa 800 200}%
\special{pa 1200 200}%
\special{fp}%
%
\special{pn 8}%
\special{pa 1200 400}%
\special{pa 1600 400}%
\special{dt 0.045}%
\special{sh 1}%
\special{pa 1600 400}%
\special{pa 1534 380}%
\special{pa 1548 400}%
\special{pa 1534 420}%
\special{pa 1600 400}%
\special{fp}%
%
\special{pn 8}%
\special{pa 1600 200}%
\special{pa 2000 200}%
\special{pa 2000 600}%
\special{pa 1600 600}%
\special{pa 1600 200}%
\special{pa 2000 200}%
\special{fp}%
%
\special{pn 8}%
\special{pa 2000 400}%
\special{pa 2400 400}%
\special{dt 0.045}%
\special{sh 1}%
\special{pa 2400 400}%
\special{pa 2334 380}%
\special{pa 2348 400}%
\special{pa 2334 420}%
\special{pa 2400 400}%
\special{fp}%
%
\special{pn 8}%
\special{pa 2400 200}%
\special{pa 2800 200}%
\special{pa 2800 600}%
\special{pa 2400 600}%
\special{pa 2400 200}%
\special{pa 2800 200}%
\special{fp}%
%
\special{pn 8}%
\special{pa 2800 400}%
\special{pa 3200 400}%
\special{fp}%
\special{sh 1}%
\special{pa 3200 400}%
\special{pa 3134 380}%
\special{pa 3148 400}%
\special{pa 3134 420}%
\special{pa 3200 400}%
\special{fp}%
\put(10.0000,-4.0000){\makebox(0,0){$\samp{h}$}}%
\put(18.0000,-4.0000){\makebox(0,0){$K(z)$}}%
\put(26.0000,-4.0000){\makebox(0,0){$\hold{h}$}}%
\put(4.0000,-3.5000){\makebox(0,0)[lb]{$u$}}%
\put(13.0000,-3.5000){\makebox(0,0)[lb]{$v$}}%
\put(21.0000,-3.5000){\makebox(0,0)[lb]{$\psi$}}%
\put(30.5000,-3.5000){\makebox(0,0)[lb]{$\hat{y}$}}%
\end{picture}%

 \caption{Digital system $\K$ consisting of ideal sampler $\samp{h}$,
 digital filter $K(z)$, and zero-order hold $\hold{h}$ with sampling period $h$.}
 \label{fig:digital-system}
\end{figure}
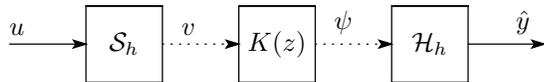

Assume that a transfer function $G(s)$ of an analog filter is given.
We suppose that $G(s)$ is a stable, real-rational, proper transfer function.
Let $g(t)$ denote the impulse response (or the inverse Laplace transform) of $G(s)$.
Our objective is to find a digital filter $K(z)$ in a digital system
\[
 \K = \hold{h}K\samp{h},
\]
shown in Fig.~\ref{fig:digital-system}
that mimics the input/output behavior of the analog filter $G(s)$.
The digital system in Fig.~\ref{fig:digital-system} includes 
an ideal sampler $\samp{h}$ and a zero-order hold $\hold{h}$ synchronized with a fixed sampling period $h>0$.
The ideal sampler $\samp{h}$ converts a continuous-time signal $u(t)$ 
to a discrete-time signal $v[n]$ as
\[
 v[n] = (\samp{h}u)[n] = u(nh),\quad n=0,1,2,\dots.
\]
The zero-order hold $\hold{h}$ produces a continuous-time signal $\hat{y}(t)$ 
from a discrete-time signal $\psi[n]$ as
\[
 \hat{y}(t) = \sum_{n=0}^\infty \psi[n]\phi(t-nh),\quad t\in[0,\infty),
\]
where $\phi(t)$ is a box function defined by
\[
 \phi(t) = \begin{cases} 1, & \text{~if~} t\in [0,h),\\ 0, & \text{~otherwise.}\end{cases}
\]

The filter $K(z)$ is designed to produce a continuous-time signal $\hat{y}$ after
the zero-order hold $\hold{h}$ that
approximates the \emph{delayed} output 
\[
 y(t)=(g\ast u)(t-l)=\int_0^{t-l}g(\tau)u(t-l-\tau)d\tau
\]
of an input $u$.
A positive delay time $l$ may improve the approximation performance when $l$ is large enough
as discussed in e.g., \cite{NagOguYam11,YamNagKha12}.
We here assume $l$ is an integer multiple of $h$, that is, $l=mh$,
where $m$ is a nonnegative integer.
Then, to avoid a trivial solution (i.e., $K(z)=0$),
we should assume some \emph{a priori} information for the inputs.
As used in \cite{NagOguYam11,YamNagKha12},
we adopt the following signal subspace of $L^2[0,\infty)$ to which the inputs belong:
\[
 FL^2 := \left\{Fw: w\in L^2[0,\infty)\right\},
\]
where $F$ is a linear system with a stable, real-rational, strictly proper transfer function $F(s)$.
This transfer function, $F(s)$, defines the \emph{analog characteristic} of the input signals
in the frequency domain.

In summary, our discretization problem is formulated as follows:
\begin{problem}
\label{prob:hinf}
Given target filter $G(s)$, analog characteristic $F(s)$,
sampling period $h$, and delay step $m$,
find a digital filter $K(z)$ that minimizes
\begin{equation}
 \begin{split}
 J
  &=\sup_{w\in L^2,\,\|w\|_2=1}\left\|\left(\ee^{-mhs}G-\hold{h}K\samp{h}\right)Fw\right\|_2\\
  &=\left\|\left(\ee^{-mhs}G-\hold{h}K\samp{h}\right)F\right\|_\infty.
 \end{split} 
 \label{eq:hinf-cost}
\end{equation}
The corresponding block diagram of the erros system
\begin{equation}
 \E := \left(\ee^{-mhs}G-\hold{h}K\samp{h}\right)F
 \label{eq:error-system}
\end{equation}
is shown in Fig.~\ref{fig:error-system}. 
\begin{figure}[tb]
 \centering
\unitlength 0.1in
\begin{picture}( 32.5000, 10.0000)(  3.0000,-12.0000)
%
\special{pn 8}%
\special{pa 1200 1008}%
\special{pa 1400 1008}%
\special{fp}%
\special{sh 1}%
\special{pa 1400 1008}%
\special{pa 1334 988}%
\special{pa 1348 1008}%
\special{pa 1334 1028}%
\special{pa 1400 1008}%
\special{fp}%
%
\special{pn 8}%
\special{pa 1400 800}%
\special{pa 1800 800}%
\special{pa 1800 1200}%
\special{pa 1400 1200}%
\special{pa 1400 800}%
\special{pa 1800 800}%
\special{fp}%
%
\special{pn 8}%
\special{pa 1800 1000}%
\special{pa 2000 1000}%
\special{dt 0.045}%
\special{sh 1}%
\special{pa 2000 1000}%
\special{pa 1934 980}%
\special{pa 1948 1000}%
\special{pa 1934 1020}%
\special{pa 2000 1000}%
\special{fp}%
%
\special{pn 8}%
\special{pa 2000 800}%
\special{pa 2400 800}%
\special{pa 2400 1200}%
\special{pa 2000 1200}%
\special{pa 2000 800}%
\special{pa 2400 800}%
\special{fp}%
%
\special{pn 8}%
\special{pa 2400 1000}%
\special{pa 2600 1000}%
\special{dt 0.045}%
\special{sh 1}%
\special{pa 2600 1000}%
\special{pa 2534 980}%
\special{pa 2548 1000}%
\special{pa 2534 1020}%
\special{pa 2600 1000}%
\special{fp}%
%
\special{pn 8}%
\special{pa 2600 800}%
\special{pa 3000 800}%
\special{pa 3000 1200}%
\special{pa 2600 1200}%
\special{pa 2600 800}%
\special{pa 3000 800}%
\special{fp}%
\put(16.0000,-10.1000){\makebox(0,0){$\samp{h}$}}%
\put(22.0000,-10.0000){\makebox(0,0){$K(z)$}}%
\put(28.0000,-10.0000){\makebox(0,0){$\hold{h}$}}%
\put(12.7000,-7.0000){\makebox(0,0){$u$}}%
\put(31.0000,-10.5000){\makebox(0,0)[lt]{$\hat{y}$}}%
%
\special{pn 8}%
\special{pa 1200 1008}%
\special{pa 1200 408}%
\special{fp}%
%
\special{pn 8}%
\special{pa 1700 200}%
\special{pa 2100 200}%
\special{pa 2100 600}%
\special{pa 1700 600}%
\special{pa 1700 200}%
\special{pa 2100 200}%
\special{fp}%
%
\special{pn 8}%
\special{pa 2300 200}%
\special{pa 2700 200}%
\special{pa 2700 600}%
\special{pa 2300 600}%
\special{pa 2300 200}%
\special{pa 2700 200}%
\special{fp}%
%
\special{pn 8}%
\special{pa 1200 408}%
\special{pa 1700 408}%
\special{fp}%
\special{sh 1}%
\special{pa 1700 408}%
\special{pa 1634 388}%
\special{pa 1648 408}%
\special{pa 1634 428}%
\special{pa 1700 408}%
\special{fp}%
%
\special{pn 8}%
\special{pa 2100 400}%
\special{pa 2300 400}%
\special{fp}%
\special{sh 1}%
\special{pa 2300 400}%
\special{pa 2234 380}%
\special{pa 2248 400}%
\special{pa 2234 420}%
\special{pa 2300 400}%
\special{fp}%
%
\special{pn 8}%
\special{pa 2700 400}%
\special{pa 3200 400}%
\special{fp}%
%
\special{pn 8}%
\special{pa 1200 700}%
\special{pa 1000 700}%
\special{fp}%
%
\special{pn 8}%
\special{ar 3200 700 50 50  0.0000000  6.2831853}%
%
\special{pn 8}%
\special{pa 3200 400}%
\special{pa 3200 650}%
\special{fp}%
\special{sh 1}%
\special{pa 3200 650}%
\special{pa 3220 584}%
\special{pa 3200 598}%
\special{pa 3180 584}%
\special{pa 3200 650}%
\special{fp}%
%
\special{pn 8}%
\special{pa 3000 1000}%
\special{pa 3200 1000}%
\special{fp}%
%
\special{pn 8}%
\special{pa 3200 1000}%
\special{pa 3200 750}%
\special{fp}%
\special{sh 1}%
\special{pa 3200 750}%
\special{pa 3180 818}%
\special{pa 3200 804}%
\special{pa 3220 818}%
\special{pa 3200 750}%
\special{fp}%
%
\special{pn 8}%
\special{pa 3250 700}%
\special{pa 3550 700}%
\special{fp}%
\special{sh 1}%
\special{pa 3550 700}%
\special{pa 3484 680}%
\special{pa 3498 700}%
\special{pa 3484 720}%
\special{pa 3550 700}%
\special{fp}%
\put(31.0000,-3.5000){\makebox(0,0)[lb]{$y$}}%
\put(34.5000,-6.5000){\makebox(0,0)[lb]{$e$}}%
%
\special{pn 8}%
\special{pa 1000 500}%
\special{pa 600 500}%
\special{pa 600 900}%
\special{pa 1000 900}%
\special{pa 1000 500}%
\special{pa 600 500}%
\special{fp}%
%
\special{pn 8}%
\special{pa 300 700}%
\special{pa 600 700}%
\special{fp}%
\special{sh 1}%
\special{pa 600 700}%
\special{pa 534 680}%
\special{pa 548 700}%
\special{pa 534 720}%
\special{pa 600 700}%
\special{fp}%
\put(8.0000,-7.0000){\makebox(0,0){$F(s)$}}%
\put(19.0000,-4.0000){\makebox(0,0){$G(s)$}}%
\put(25.0000,-4.0000){\makebox(0,0){$e^{-mhs}$}}%
\put(3.0000,-6.5000){\makebox(0,0)[lb]{$w$}}%
\put(32.5000,-6.5000){\makebox(0,0)[lb]{$+$}}%
\put(32.5000,-7.5000){\makebox(0,0)[lt]{$-$}}%
\end{picture}%

 \caption{Error system $\E$.}
 \label{fig:error-system}
\end{figure}
\end{problem}

\section{STEP-INVARIANT AND BILINEAR METHODS}
\label{sec:conventional}
Before solving Problem \ref{prob:hinf},
we here briefly review two conventional discretization methods,
namely step-invariant transformation and bilinear transformation
\cite{CheFra}.
\subsection{Step-invariant transformation}
The step-invariant transformation $K_\dd$ of a continuous-time system $G$
is defined by
\[
 K_\dd := \samp{h}G\hold{h}.
\]
When the continuous-time input $u$ applied to $G$ is the step function:
\[
 u(t) = \one(t) := \begin{cases} 1, & \text{~if~} t\geq 0,\\ 0, & \text{~if~} t<0, \end{cases}
\]
then the output $y$ and the approximation $\hat{y}$ processed by the digital system
shown in Fig.~\ref{fig:digital-system} with $K=K_\dd$ are equal
on the sampling instants, $t=0,h,2h,\dots$, that is,
\[
 y(nh) = \hat{y}(nh),\quad n=0,1,2,\dots.
\]
In other words, we have
\[
 K_\dd \samp{h} \one = \samp{h}G\one.
\]
The term ``step-invariant'' is derived from this property.

If $G(s)$ has a state-space representation
\begin{equation}
 G(s) = \abcd{A}{B}{C}{D} = C(sI-A)^{-1}B + D,
 \label{eq:abcd}
\end{equation}
then the step-invariant transformation $K_\dd(z)$ has the following state-space representation
\cite[Theorem 3.1.1]{CheFra}:
\begin{equation}
 K_\dd(z) = \abcd{A_\dd}{B_\dd}{C}{D} = C(zI-A_\dd)^{-1}B_\dd + D,
 \label{eq:step-invariant}
\end{equation}
where
\[
 A_\dd := \ee^{Ah}, \quad B_\dd = \int_0^h \ee^{At}B\dd t.
\]
We denote the above transformation by $\ctod(G,h)$.
\subsection{Bilinear transformation}
Another well-known discretization method is \emph{bilinear transformation},
also known as \emph{Tustin's method}.
This is based on the trapezoidal rule for approximating the definite integral.
The bilinear transformation $K_\bt(z)$ of a continuous-time transfer function $G(s)$
is given by
\[
 K_\bt(z) = G\left(\frac{2}{h}\frac{z-1}{z+1}\right).
\]
The mapping from $s$ to $z$ is given by
\[
 s \mapsto z = \frac{1+(h/2)s}{1-(h/2)s},
\]
and this maps the open left half plane into the open unit circle,
and hence stability and minimum-phase property are preserved under bilinear transformation.

A state-space representation of bilinear transformation $K_\bt$ can be
obtained as follows.
Assume that $G$ has a state-space representation
given in \eqref{eq:abcd},
then $K_\bt$ has the following state-space representation
\cite[Sec.~3.4]{CheFra}:
\[
 K_\bt(z) = \abcd{A_\bt}{B_\bt}{C_\bt}{D_\bt} = C_\bt(zI-A_\bt)^{-1}B_\bt + D_\bt,
\]
where
\begin{equation}
 \begin{split}
  A_\bt &:= \left(I-\frac{h}{2}A\right)^{-1}\left(I+\frac{h}{2}A\right),\\
  B_\bt &:= \frac{h}{2}\left(I-\frac{h}{2}A\right)^{-1}B,\\
  C_\bt &:= C(I+A_\bt),\\
  D_\bt &:= D + CB_\bt.
 \end{split}
 \label{eq:bt-abcd}
\end{equation}

The frequency response of the bilinear transformation $K_\bt$ at frequency $\omega$ (rad/sec)
is given by
\[
 K_\bt\left(\ee^{\jj\omega h}\right) 
 = G\left(\frac{2}{h}\frac{\ee^{\jj\omega h}-1}{\ee^{\jj\omega h}+1}\right)
 = G\left(\jj\cdot\frac{h}{2}\tan \frac{\omega h}{2}\right).
\]
It follows that at $\omega=0$ (rad/sec), the frequency responses $K_\bt(\ee^{\jj\omega h})$ and $G(\jj\omega)$
coincide, and hence there is no error at DC (direct current).
If a zero error is preferred instead at another frequency, say $\omega=\omega_0$ (rad/sec),
\emph{frequency prewarping} may be used for this purpose.
The bilinear transformation with frequency prewarping
is given by
\begin{equation}
 \begin{split}
  K_{\bt,\omega_0}(z) &= G\left(c(\omega_0)\cdot\frac{z-1}{z+1}\right),\\
  c(\omega_0) &:= \omega_0\left(\tan\frac{\omega_0 h}{2}\right)^{-1}.
 \end{split}
 \label{eq:bilinear-prewarp}
\end{equation}
It is easily proved that $K_{\bt,\omega_0}(\ee^{\jj\omega_0 h})=G(\jj\omega_0)$,
that is, there is no error at $\omega=\omega_0$ (rad/sec).
A state-space representation of $K_{\bt,\omega_0}$ can be obtained 
by using the same formula
\eqref{eq:bt-abcd} with $1/c(\omega_0)$ instead of $h/2$.

\section{$H^\infty$-OPTIMAL DISCRETIZATION}
\label{sec:H-inf}

In this section, we give a design formula to numerically compute the $H^\infty$-optimal filter
of Problem \ref{prob:hinf} via \emph{fast sample/hold} approximation
\cite{KelAnd92,CheFra,YamMadAnd99,YamAndNag02}.
This method approximates
a continuous-time signal of interest by
a piecewise constant signal,
which is obtained by a fast sampler followed by a fast hold
with period $h/N$ for some positive integer $N$.
The convergence of the approximation
is shown in \cite{YamMadAnd99,YamAndNag02}.

Before proceeding,
we define \emph{discrete-time lifting} $\dlift{N}$ and its inverse 
$\idlift{N}$ for a discrete-time signal
as
\begin{equation}
\begin{split}
    \dlift{N} &:= (\ds{N})
	\left[\begin{array}{cccc}
	  1 & z & \cdots & z^{N-1}
	\end{array}\right]^T,\\
    \idlift{N} &:=
 	\left[\begin{array}{cccc}
	  1 & z^{-1} & \cdots & z^{-N+1}
	\end{array}\right](\us{N}),
\end{split}
\label{eq:dlift}
\end{equation}
where $\ds{N}$ and $\us{N}$ are respectively
a downsampler and an upsampler
\cite{Vai} defined as
\[
 \begin{split}
  \us{N} &: \bigl\{x[k]\bigr\}_{k=0}^\infty \mapsto \bigl\{x[0],\underbrace{0,\dots,0}_{N-1},x[1],0,\dots\bigr\},\\
  \ds{N} &: \bigl\{x[k]\bigr\}_{k=0}^\infty \mapsto \bigl\{x[0],x[N],x[2N],\dots\bigr\}.
 \end{split}
\]
The discrete-time lifting for a discrete-time \emph{system} is then
defined as
\[
 \begin{split}
  &\lift\left(\abcd{A}{B}{C}{D},N\right)\\
  &\quad := \dlift{N}\abcd{A}{B}{C}{D}\idlift{N}\\
  &\quad = \left[\begin{array}{c|cccc}
	 A^N & A^{N-1}B & A^{N-2}B & \ldots & B\\\hline
	 C & D & 0 & \ldots & 0\\
	 CA & CB & D & \ddots & \vdots\\
	 \vdots & \vdots & \vdots & \ddots & 0\\
	 CA^{N-1} & CA^{N-2}B& CA^{N-3}B & \ldots & D
    \end{array}\right].
 \end{split}    
\]

By using the discrete-time lifting, we obtain the fast sample/hold approximation $E_N$
of the sampled-data error system $\E$ given in \eqref{eq:error-system}
as
\begin{equation}
 E_N(z) = \bigl(z^{-mN}G_N(z) - H_NK(z)S_N\bigr)F_N(z),
 \label{eq:EN}
\end{equation}
where
\begin{equation}
 \begin{split}
  G_N(z) &= \lift\bigl(\ctod(G(s),h/N\bigr),N),\\
  F_N(z) &= \lift\bigl(\ctod(F(s),h/N),N\bigr),\\
  H_N &= [\underbrace{1,\dots,1}_{N}]^\top,\quad
  S_N = [1,\underbrace{0,\dots,0}_{N-1}].
 \end{split}
 \label{eq:formula}
\end{equation}

The optimal discretization problem formulated in Problem \ref{prob:hinf}
is then approximated by a standard discrete-time $H^\infty$ optimization
problem:
we find the optimal $K(z)$ that minimizes the \emph{discrete-time} $H^\infty$ norm
of $E_N(z)$, that is,
\[
 \min_{K} \|E_N\|_\infty = \min_{K} \left\|\bigl(z^{-mN}G_N-H_NKS_N\bigr)\right\|_\infty.
\]
The minimizer can be effectively computed by a standard numerical computation software
such as MATLAB.

Moreover, a convergence result is obtained as follows.
\begin{theorem}
For each fixed $K(z)$ and for each $\omega \in [0,2\pi/h)$, the frequency response
satisfies
\[
 \|E_N(\ee^{\jj\omega h})\| \rightarrow \|\widetilde{\E}(\ee^{\jj\omega h})\|,~ \text{~as~} N\rightarrow\infty,
\]
where $\widetilde{\E}$ is the lifted system of $\E$.
The convergence is uniform with respect to $\omega\in[0,2\pi/h)$
and with $K$ in a compact set of stable filters.
\end{theorem}
{\bf Proof.}
This is a direct consequence of results in \cite{YamMadAnd99,YamAndNag02}.
\hfill $\Box$

The theorem guarantees that for sufficiently large $N$ 
the error becomes small enough with a filter
$K(z)$ obtained by the fast sample/hold approximation.

\begin{remark}
If an FIR (finite impulse response) filter is preferred for filter $K(z)$,
it can be obtained via LMI (linear matrix inequality) optimization.
See \cite{YamAndNagKoy03,Nag11} for details.
\end{remark}

\section{EXTENSION TO MULTIRATE SYSTEMS}
\label{sec:multirate}

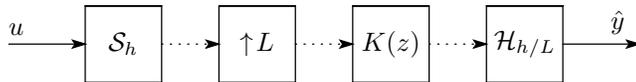
\begin{figure}[tb]
 \centering
\unitlength 0.1in
\begin{picture}( 33.0000,  4.0000)(  4.0000, -6.0000)
%
\special{pn 8}%
\special{pa 400 400}%
\special{pa 800 400}%
\special{fp}%
\special{sh 1}%
\special{pa 800 400}%
\special{pa 734 380}%
\special{pa 748 400}%
\special{pa 734 420}%
\special{pa 800 400}%
\special{fp}%
%
\special{pn 8}%
\special{pa 800 200}%
\special{pa 1200 200}%
\special{pa 1200 600}%
\special{pa 800 600}%
\special{pa 800 200}%
\special{pa 1200 200}%
\special{fp}%
%
\special{pn 8}%
\special{pa 1200 400}%
\special{pa 1500 400}%
\special{dt 0.045}%
\special{sh 1}%
\special{pa 1500 400}%
\special{pa 1434 380}%
\special{pa 1448 400}%
\special{pa 1434 420}%
\special{pa 1500 400}%
\special{fp}%
%
\special{pn 8}%
\special{pa 1500 200}%
\special{pa 1900 200}%
\special{pa 1900 600}%
\special{pa 1500 600}%
\special{pa 1500 200}%
\special{pa 1900 200}%
\special{fp}%
%
\special{pn 8}%
\special{pa 1900 400}%
\special{pa 2200 400}%
\special{dt 0.045}%
\special{sh 1}%
\special{pa 2200 400}%
\special{pa 2134 380}%
\special{pa 2148 400}%
\special{pa 2134 420}%
\special{pa 2200 400}%
\special{fp}%
%
\special{pn 8}%
\special{pa 2200 200}%
\special{pa 2600 200}%
\special{pa 2600 600}%
\special{pa 2200 600}%
\special{pa 2200 200}%
\special{pa 2600 200}%
\special{fp}%
%
\special{pn 8}%
\special{pa 2600 400}%
\special{pa 2900 400}%
\special{dt 0.045}%
\special{sh 1}%
\special{pa 2900 400}%
\special{pa 2834 380}%
\special{pa 2848 400}%
\special{pa 2834 420}%
\special{pa 2900 400}%
\special{fp}%
%
\special{pn 8}%
\special{pa 2900 200}%
\special{pa 3300 200}%
\special{pa 3300 600}%
\special{pa 2900 600}%
\special{pa 2900 200}%
\special{pa 3300 200}%
\special{fp}%
%
\special{pn 8}%
\special{pa 3300 400}%
\special{pa 3700 400}%
\special{fp}%
\special{sh 1}%
\special{pa 3700 400}%
\special{pa 3634 380}%
\special{pa 3648 400}%
\special{pa 3634 420}%
\special{pa 3700 400}%
\special{fp}%
\put(17.0000,-4.0000){\makebox(0,0){$\us{L}$}}%
\put(31.0000,-4.0000){\makebox(0,0){$\hold{h/L}$}}%
\put(24.0000,-4.0000){\makebox(0,0){$K(z)$}}%
\put(10.0000,-4.0000){\makebox(0,0){$\samp{h}$}}%
\put(4.0000,-3.5000){\makebox(0,0)[lb]{$u$}}%
\put(35.5000,-3.5000){\makebox(0,0)[lb]{$\hat{y}$}}%
\end{picture}%

 \caption{Multirate system $\K_\mr$ consisting of ideal sampler $\samp{h}$,
 upsampler $\us{L}$, 
 digital filter $K(z)$, and fast hold $\hold{h/L}$.}
 \label{fig:multirate-system}
\end{figure}

\begin{figure}[tb]
 \centering
\unitlength 0.1in
\begin{picture}( 33.5000, 12.0000)(  3.0000,-14.0000)
%
\special{pn 8}%
\special{pa 1050 1198}%
\special{pa 1200 1198}%
\special{fp}%
\special{sh 1}%
\special{pa 1200 1198}%
\special{pa 1134 1178}%
\special{pa 1148 1198}%
\special{pa 1134 1218}%
\special{pa 1200 1198}%
\special{fp}%
%
\special{pn 8}%
\special{pa 1200 1000}%
\special{pa 1600 1000}%
\special{pa 1600 1400}%
\special{pa 1200 1400}%
\special{pa 1200 1000}%
\special{pa 1600 1000}%
\special{fp}%
%
\special{pn 8}%
\special{pa 1600 1200}%
\special{pa 1750 1200}%
\special{dt 0.045}%
\special{sh 1}%
\special{pa 1750 1200}%
\special{pa 1684 1180}%
\special{pa 1698 1200}%
\special{pa 1684 1220}%
\special{pa 1750 1200}%
\special{fp}%
%
\special{pn 8}%
\special{pa 1750 1000}%
\special{pa 2150 1000}%
\special{pa 2150 1400}%
\special{pa 1750 1400}%
\special{pa 1750 1000}%
\special{pa 2150 1000}%
\special{fp}%
%
\special{pn 8}%
\special{pa 2150 1200}%
\special{pa 2300 1200}%
\special{dt 0.045}%
\special{sh 1}%
\special{pa 2300 1200}%
\special{pa 2234 1180}%
\special{pa 2248 1200}%
\special{pa 2234 1220}%
\special{pa 2300 1200}%
\special{fp}%
%
\special{pn 8}%
\special{pa 2300 1000}%
\special{pa 2700 1000}%
\special{pa 2700 1400}%
\special{pa 2300 1400}%
\special{pa 2300 1000}%
\special{pa 2700 1000}%
\special{fp}%
%
\special{pn 8}%
\special{pa 2700 1200}%
\special{pa 2850 1200}%
\special{dt 0.045}%
\special{sh 1}%
\special{pa 2850 1200}%
\special{pa 2784 1180}%
\special{pa 2798 1200}%
\special{pa 2784 1220}%
\special{pa 2850 1200}%
\special{fp}%
%
\special{pn 8}%
\special{pa 2850 1000}%
\special{pa 3250 1000}%
\special{pa 3250 1400}%
\special{pa 2850 1400}%
\special{pa 2850 1000}%
\special{pa 3250 1000}%
\special{fp}%
\put(19.5000,-12.0000){\makebox(0,0){$\us{L}$}}%
\put(30.5000,-12.0000){\makebox(0,0){$\hold{h/L}$}}%
\put(25.0000,-12.0000){\makebox(0,0){$K(z)$}}%
\put(14.0000,-12.0000){\makebox(0,0){$\samp{h}$}}%
%
\special{pn 8}%
\special{pa 1050 1198}%
\special{pa 1050 398}%
\special{fp}%
%
\special{pn 8}%
\special{pa 1050 398}%
\special{pa 1800 398}%
\special{fp}%
\special{sh 1}%
\special{pa 1800 398}%
\special{pa 1734 378}%
\special{pa 1748 398}%
\special{pa 1734 418}%
\special{pa 1800 398}%
\special{fp}%
%
\special{pn 8}%
\special{pa 1800 200}%
\special{pa 2200 200}%
\special{pa 2200 600}%
\special{pa 1800 600}%
\special{pa 1800 200}%
\special{pa 2200 200}%
\special{fp}%
%
\special{pn 8}%
\special{pa 2200 400}%
\special{pa 2400 400}%
\special{fp}%
\special{sh 1}%
\special{pa 2400 400}%
\special{pa 2334 380}%
\special{pa 2348 400}%
\special{pa 2334 420}%
\special{pa 2400 400}%
\special{fp}%
%
\special{pn 8}%
\special{pa 2400 200}%
\special{pa 2800 200}%
\special{pa 2800 600}%
\special{pa 2400 600}%
\special{pa 2400 200}%
\special{pa 2800 200}%
\special{fp}%
%
\special{pn 8}%
\special{pa 2800 400}%
\special{pa 3400 400}%
\special{fp}%
%
\special{pn 8}%
\special{pa 3400 400}%
\special{pa 3400 750}%
\special{fp}%
\special{sh 1}%
\special{pa 3400 750}%
\special{pa 3420 684}%
\special{pa 3400 698}%
\special{pa 3380 684}%
\special{pa 3400 750}%
\special{fp}%
%
\special{pn 8}%
\special{pa 3250 1200}%
\special{pa 3400 1200}%
\special{fp}%
%
\special{pn 8}%
\special{pa 3400 1200}%
\special{pa 3400 850}%
\special{fp}%
\special{sh 1}%
\special{pa 3400 850}%
\special{pa 3380 918}%
\special{pa 3400 904}%
\special{pa 3420 918}%
\special{pa 3400 850}%
\special{fp}%
%
\special{pn 8}%
\special{ar 3400 800 50 50  0.0000000  6.2831853}%
%
\special{pn 8}%
\special{pa 3450 800}%
\special{pa 3650 800}%
\special{fp}%
\special{sh 1}%
\special{pa 3650 800}%
\special{pa 3584 780}%
\special{pa 3598 800}%
\special{pa 3584 820}%
\special{pa 3650 800}%
\special{fp}%
\put(36.0000,-7.5000){\makebox(0,0)[lb]{$e$}}%
\put(34.5000,-7.5000){\makebox(0,0)[lb]{$+$}}%
\put(34.5000,-8.5000){\makebox(0,0)[lt]{$-$}}%
\put(34.0000,-3.5000){\makebox(0,0)[rb]{$y$}}%
\put(34.5000,-12.5000){\makebox(0,0)[rt]{$\hat{y}$}}%
%
\special{pn 8}%
\special{pa 900 798}%
\special{pa 1050 798}%
\special{fp}%
%
\special{pn 8}%
\special{pa 900 598}%
\special{pa 500 598}%
\special{pa 500 998}%
\special{pa 900 998}%
\special{pa 900 598}%
\special{pa 500 598}%
\special{fp}%
%
\special{pn 8}%
\special{pa 300 790}%
\special{pa 500 790}%
\special{fp}%
\special{sh 1}%
\special{pa 500 790}%
\special{pa 434 770}%
\special{pa 448 790}%
\special{pa 434 810}%
\special{pa 500 790}%
\special{fp}%
\put(3.0000,-7.4700){\makebox(0,0)[lb]{$w$}}%
\put(11.5000,-7.9700){\makebox(0,0){$u$}}%
\put(7.0000,-7.9700){\makebox(0,0){$F(s)$}}%
\put(20.0000,-4.0000){\makebox(0,0){$G(s)$}}%
\put(26.0000,-4.0000){\makebox(0,0){$e^{-mhs}$}}%
\end{picture}%

 \caption{Multirate error system $\E_\mr$.}
 \label{fig:multirate-error-system}
\end{figure}
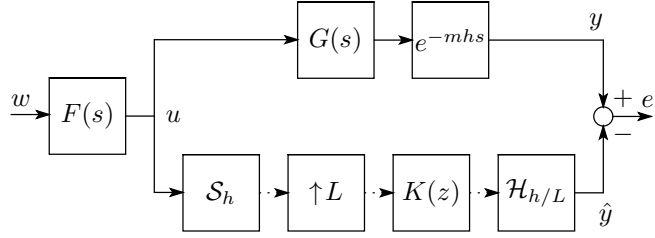

In this section, we extend the result in the previous section
to multirate systems.
If one uses a fast hold device with period
$h/L$ ($L$ is an integer greater than or equal to $2$) instead of $\hold{h}$,
the performance may be further improved.
This observation suggests us to use a multirate signal processing system
\[
 \K_\mr := \hold{h/L}K(z)(\us{L})\samp{h},
\]
shown in Fig.~\ref{fig:multirate-system}.
The objective here is to find a digital filter $K(z)$
that minimizes the $H^\infty$ norm of the multirate
error system $\E_\mr$ shown in Fig.~\ref{fig:multirate-error-system}
between the delayed target system
$G(s)\ee^{-mhs}$ and the multirate system $\K_\mr$
with analog characteristic $F(s)$.
More precisely, we solve the following problem.
\begin{problem}
\label{prob:multirate-hinf}
Given target filter $G(s)$, analog characteristic $F(s)$,
sampling period $h$, delay step $m$,
and upsampling ratio $L$,
find a digital filter $K(z)$ that minimizes
\begin{equation}
 \begin{split}
 J_\mr
  &=\sup_{w\in L^2,\,\|w\|_2=1}\left\|\left(\ee^{-mhs}G-\hold{h/L}K(\us{L})\samp{h}\right)Fw\right\|_2\\
  &=\left\|\left(\ee^{-mhs}G-\hold{h/L}K(\us{L})\samp{h}\right)F\right\|_\infty.
 \end{split} 
 \label{eq:hinf-cost-multirate}
\end{equation}
See the corresponding block diagram of the error system $\E_\mr$
shown in Fig.~\ref{fig:multirate-error-system}.
\end{problem}

We use the method of fast sample/hold approximation
to compute the optimal filter $K(z)$.
Assume that $N=Ll$ for some positive integer $p$,
then the approximated discrete-time system for $\E_\mr$ is given by
\begin{equation}
 E_{\mr,N}(z) = \bigl(G_N(z)z^{-mN} - \widetilde{H}_{N}\widetilde{K}(z)S_N\bigr)F_N(z),
\end{equation}
where $G_N(z)$, $F_N(z)$ and $S_N$ are given in \eqref{eq:formula}, and
\[
 \begin{split}
  \widetilde{H}_{N} &= \blkdiag\{\underbrace{\one_p,\one_p,\dots,\one_p}_L\},~
  \one_p = [\underbrace{1,\dots,1}_{p}]^\top,\\
  \widetilde{K}(z) &= \lift\bigl(K(z),L\bigr)[1,\underbrace{0,\dots,0}_{L-1}]^\top.
 \end{split}
\]

The optimal discretization problem with a multirate system
formulated in Problem \ref{prob:multirate-hinf}
is then approximated by a standard discrete-time
$H^\infty$ optimization.
The minimizer $\widetilde{K}(z)$ can be numerically computed by
e.g. MATLAB.
Then once the filter $\widetilde{K}(z)$ is obtained,
the filter $K(z)$ in Fig.~\ref{fig:multirate-system}
is given by the following formula:
\[
 K(z) = \bigl[1, z^{-1},\dots,z^{-L+1}\bigr]\widetilde{K}(z^L).
\]

We have the following convergence theorem:
\begin{theorem}
For each fixed $\widetilde{K}(z)$ and for each $\omega \in [0,2\pi/h)$, 
the frequency response
satisfies
\[
 \|E_{\mr,N}(\ee^{\jj\omega h})\| \rightarrow \|\widetilde{\E}_\mr(\ee^{\jj\omega h})\|,~ \text{~as~} N\rightarrow\infty,
\]
where $\widetilde{\E}_\mr$ is the lifted system of $\E_\mr$.
The convergence is uniform with respect to $\omega\in[0,2\pi/h)$
and with $\widetilde{K}$ in a compact set of stable filters.
\end{theorem}
{\bf Proof.}
The proof is almost the same as in \cite[Theorem 1]{YamNagKha12}.

\hfill $\Box$

See Section \ref{sec:examples} for a simulation result,
which shows that as upsampling ratio $L$ increases,
the $H^\infty$ norm of the error system $\E$ decreases.

\section{CONTROLLER DISCRETIZATION}
\label{sec:feedback}
\begin{figure}[tb]
 \centering
\unitlength 0.1in
\begin{picture}( 10.0000, 10.0000)(  3.0000,-12.0000)
%
\special{pn 8}%
\special{pa 600 200}%
\special{pa 1000 200}%
\special{pa 1000 600}%
\special{pa 600 600}%
\special{pa 600 200}%
\special{pa 1000 200}%
\special{fp}%
%
\special{pn 8}%
\special{pa 600 800}%
\special{pa 1000 800}%
\special{pa 1000 1200}%
\special{pa 600 1200}%
\special{pa 600 800}%
\special{pa 1000 800}%
\special{fp}%
%
\special{pn 8}%
\special{pa 1000 400}%
\special{pa 1300 400}%
\special{fp}%
%
\special{pn 8}%
\special{pa 1300 400}%
\special{pa 1300 1000}%
\special{fp}%
%
\special{pn 8}%
\special{pa 1300 1000}%
\special{pa 1000 1000}%
\special{fp}%
\special{sh 1}%
\special{pa 1000 1000}%
\special{pa 1068 1020}%
\special{pa 1054 1000}%
\special{pa 1068 980}%
\special{pa 1000 1000}%
\special{fp}%
%
\special{pn 8}%
\special{pa 600 1000}%
\special{pa 300 1000}%
\special{fp}%
%
\special{pn 8}%
\special{pa 300 1000}%
\special{pa 300 400}%
\special{fp}%
%
\special{pn 8}%
\special{pa 300 400}%
\special{pa 600 400}%
\special{fp}%
\special{sh 1}%
\special{pa 600 400}%
\special{pa 534 380}%
\special{pa 548 400}%
\special{pa 534 420}%
\special{pa 600 400}%
\special{fp}%
\put(8.0000,-4.0000){\makebox(0,0){$F(s)$}}%
\put(8.0000,-10.0000){\makebox(0,0){$G(s)$}}%
\end{picture}%

 \caption{Feedback control system.}
 \label{fig:feedback}
\end{figure}
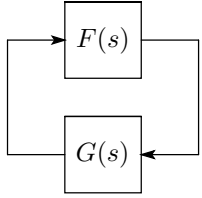

\begin{figure}[tb]
 \centering
\unitlength 0.1in
\begin{picture}( 20.0000, 10.0000)(  4.0000,-12.0000)
%
\special{pn 8}%
\special{pa 1200 200}%
\special{pa 1600 200}%
\special{pa 1600 600}%
\special{pa 1200 600}%
\special{pa 1200 200}%
\special{pa 1600 200}%
\special{fp}%
%
\special{pn 8}%
\special{pa 600 800}%
\special{pa 1000 800}%
\special{pa 1000 1200}%
\special{pa 600 1200}%
\special{pa 600 800}%
\special{pa 1000 800}%
\special{fp}%
\put(14.0000,-4.0000){\makebox(0,0){$F(s)$}}%
\put(8.0000,-10.0000){\makebox(0,0){$\hold{h}$}}%
%
\special{pn 8}%
\special{pa 1200 800}%
\special{pa 1600 800}%
\special{pa 1600 1200}%
\special{pa 1200 1200}%
\special{pa 1200 800}%
\special{pa 1600 800}%
\special{fp}%
%
\special{pn 8}%
\special{pa 1800 800}%
\special{pa 2200 800}%
\special{pa 2200 1200}%
\special{pa 1800 1200}%
\special{pa 1800 800}%
\special{pa 2200 800}%
\special{fp}%
\put(14.0000,-10.0000){\makebox(0,0){$K(z)$}}%
\put(20.0000,-10.0000){\makebox(0,0){$\samp{h}$}}%
%
\special{pn 8}%
\special{pa 1600 400}%
\special{pa 2400 400}%
\special{fp}%
\special{pa 2400 400}%
\special{pa 2400 1000}%
\special{fp}%
%
\special{pn 8}%
\special{pa 2400 1000}%
\special{pa 2200 1000}%
\special{fp}%
\special{sh 1}%
\special{pa 2200 1000}%
\special{pa 2268 1020}%
\special{pa 2254 1000}%
\special{pa 2268 980}%
\special{pa 2200 1000}%
\special{fp}%
%
\special{pn 8}%
\special{pa 600 1000}%
\special{pa 400 1000}%
\special{fp}%
\special{pa 400 1000}%
\special{pa 400 400}%
\special{fp}%
%
\special{pn 8}%
\special{pa 400 400}%
\special{pa 1200 400}%
\special{fp}%
\special{sh 1}%
\special{pa 1200 400}%
\special{pa 1134 380}%
\special{pa 1148 400}%
\special{pa 1134 420}%
\special{pa 1200 400}%
\special{fp}%
%
\special{pn 8}%
\special{pa 1800 1000}%
\special{pa 1600 1000}%
\special{dt 0.045}%
\special{sh 1}%
\special{pa 1600 1000}%
\special{pa 1668 1020}%
\special{pa 1654 1000}%
\special{pa 1668 980}%
\special{pa 1600 1000}%
\special{fp}%
%
\special{pn 8}%
\special{pa 1200 1000}%
\special{pa 1000 1000}%
\special{dt 0.045}%
\special{sh 1}%
\special{pa 1000 1000}%
\special{pa 1068 1020}%
\special{pa 1054 1000}%
\special{pa 1068 980}%
\special{pa 1000 1000}%
\special{fp}%
\end{picture}%

 \caption{Sampled-data feedback control system.}
 \label{fig:sd-feedback}
\end{figure}
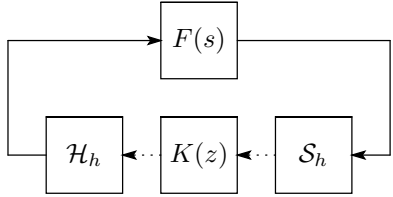

We here consider feedback controller discretization.
Let us consider the feedback system consisting of two
linear time-invariant systems, $F(s)$ and $G(s)$,
shown in Fig.~\ref{fig:feedback}.
We assume that $F(s)$ is a plant model that is
a stable, real-rational, strictly proper transfer function,
and $G(s)$ is a stable, real-rational, proper controller.
We also assume that they satisfy 
\[
\|GF\|_\infty < 1.
\]
Then, by the small-gain theorem \cite{ZhoDoyGlo},
the feedback system is stable.
Then, to implement continuous-time controller $G(s)$
in a digital system,
we discretize it by using the $H^\infty$-optimal discretization discussed above.
Suppose that we obtain an $H^\infty$-optimal digital filter $K(z)$ with no delay ($m=0$).
Then we have
\[
 \begin{split}
  \|\hold{h}K\samp{h}F\|_\infty &= \|(G-\hold{h}K\samp{h})F-GF\|_\infty\\
  &\leq \|\E\|_\infty + \|GF\|_\infty.
 \end{split}
\]
It follows that by the small-gain theorem,
the sampled-data feedback control system shown in Fig.~\ref{fig:sd-feedback}
is stable if 
$\|\E\|_\infty + \|GF\|_\infty<1$,
or equivalently
\begin{equation}
 \|\E\|_\infty < 1-\|GF\|_\infty.
 \label{eq:smallgain}
\end{equation}

In summary, if we can sufficiently decrease the $H^\infty$ norm of the sampled-data error system $\E$
via the $H^\infty$-optimal discretization to satisfy \eqref{eq:smallgain}, 
then the stability of the feedback control system is preserved
under discretization.
The multirate system proposed in Section \ref{sec:multirate} can be also used for 
stability-preserving controller discretization.

\section{NUMERICAL EXAMPLES}
\label{sec:examples}

\begin{figure}[tb]
\centering
\includegraphics[width=\linewidth]{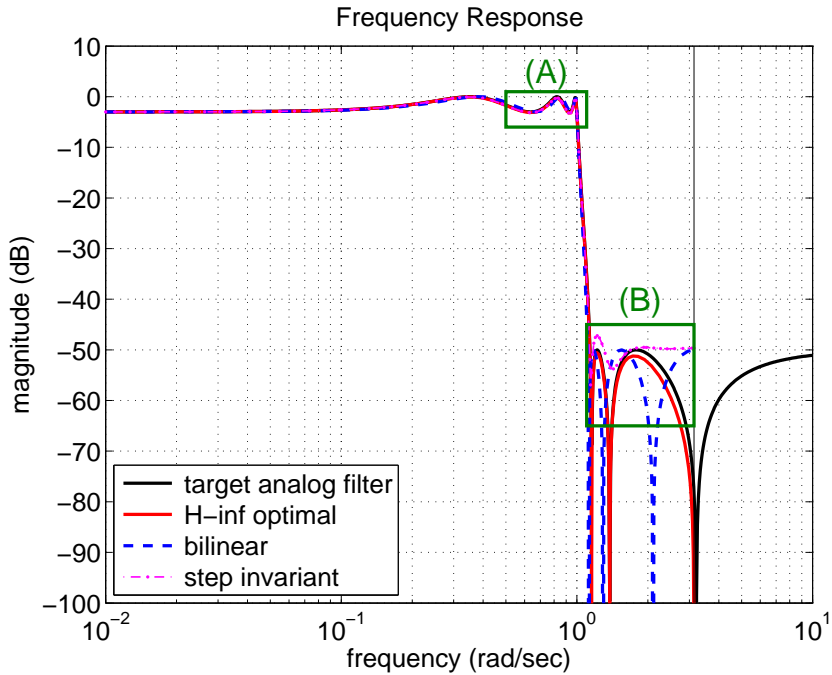}
\caption{Frequency response of filters: target analog filter $G(s)$ (solid black),
sampled-data $H^\infty$-optimal $K_\sd(z)$ (solid red),
bilinear transformation $K_{\bt,\omega_0}(z)$ with frequency prewarping at
$\omega_0=1$ (rad/sec) (dashed blue),
and step-invariant transformation (dash-dotted magenta).}
\label{fig:freq-resp}
\end{figure}

\begin{figure}[tb]
\centering
\includegraphics[width=\linewidth]{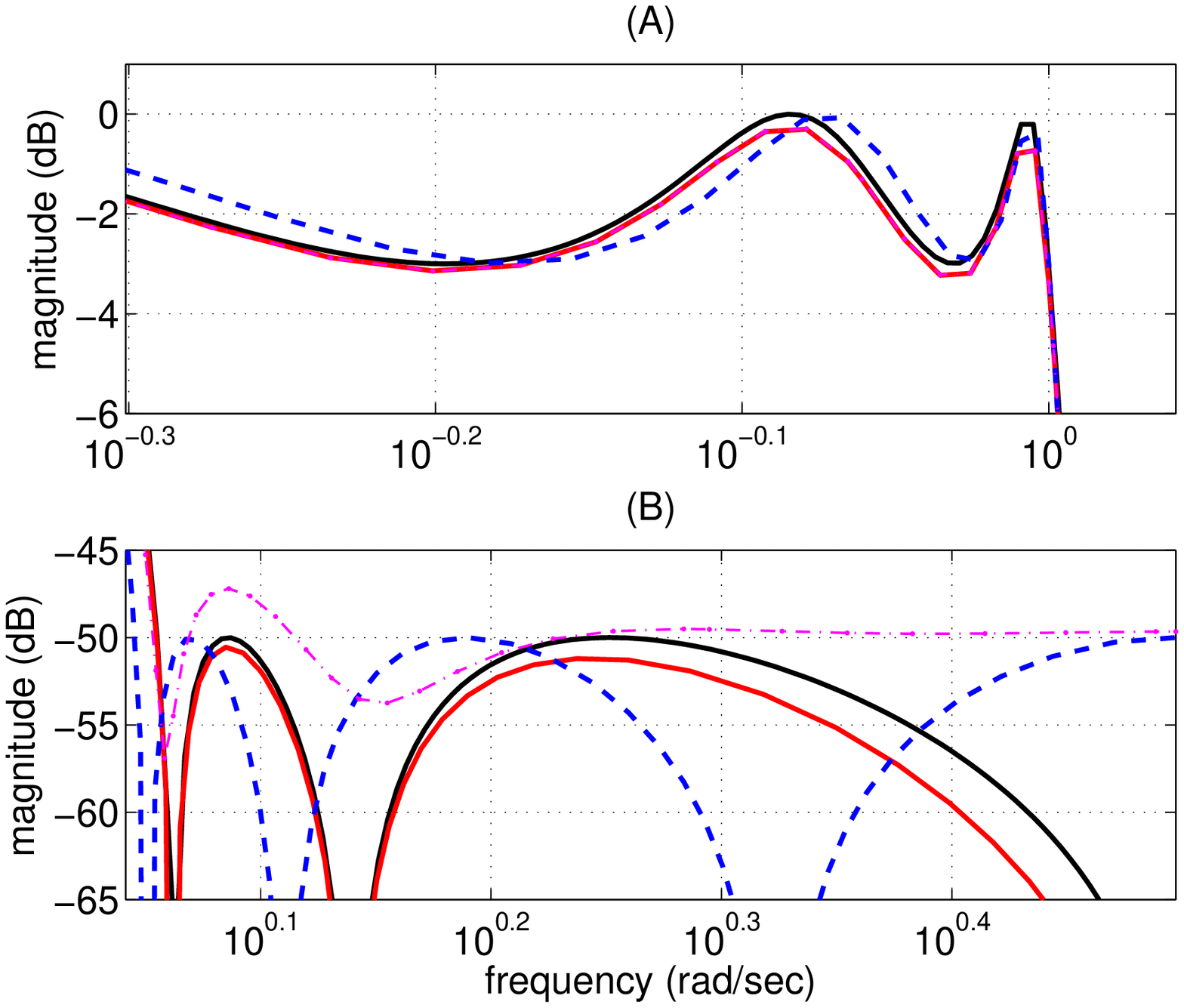}
\caption{Enlarged plots of Fig.~\ref{fig:freq-resp} in
passband (top) and stopband (bottom).}
\label{fig:freq-resp-e}
\end{figure}

Here we present numerical examples to illustrate the effectiveness of the proposed method.
The target analog filter is given as a $6$-th order elliptic filter
with $3$ (dB) passband peak-to-peak ripple, 
$-50$ (dB) stopband attenuation, 
and 1 (rad/sec) cut-off frequency.
The filter is computed with MATLAB command
\verb=ellip(6,3,50,1,'s')=
whose transfer function is given by
\[
 \begin{split}
  G(s)=&\frac{0.0031623 (s^2 + 1.33) (s^2 + 1.899)}{(s^2 + 0.3705s + 0.1681) (s^2 + 0.1596s + 0.7062)}\\
  &\quad \times \frac{ (s^2 + 10.31)}{ (s^2 + 0.03557s + 0.9805)}.
 \end{split}
\]
We use the following analog characteristic
\[
 F(s) = \frac{1}{(s+1)^3}.
\]
We set sampling period $h=1$, delay step $m=4$, and 
the ratio
for fast sample/hold approximation $N=12$.
With these parameters, we design the $H^\infty$-optimal filter, denoted by $K_\sd(z)$
that minimizes the cost function $J$ in \eqref{eq:hinf-cost}.
We also design the step-invariant transformation $K_\dd(z)$ given in \eqref{eq:step-invariant}
and the bilinear transformation $K_{\bt,\omega_0}(z)$ with frequency prewarping at $\omega_0=1$ (rad/sec)
given in \eqref{eq:bilinear-prewarp}.
Fig.~\ref{fig:freq-resp} shows the frequency response of these filters.
Enlarged plots of the frequency response in passband and stopband are also shown 
in Fig.~\ref{fig:freq-resp-e}
These plots show that the proposed sampled-data $H^\infty$-optimal discretization
$K_\sd(z)$ shows the best approximation among the filters.
The step-invariant transformation $K_\dd(z)$ is almost the same as 
$K_\sd(z)$ in passband,
while its stopband response is quite different from that of the target filter $G(s)$.
The bilinear transformation $K_{\bt,\omega_0}(z)$ shows the best performance
around the cut-off frequency $\omega_0=1$ (rad/sec)
at the cost of deterioration of performance at the other frequencies.
In summary, the sampled-data $H^\infty$ discretization outperforms
conventional discretization methods
in particular in high-frequency range.

\begin{figure}[tb]
\centering
\includegraphics[width=\linewidth]{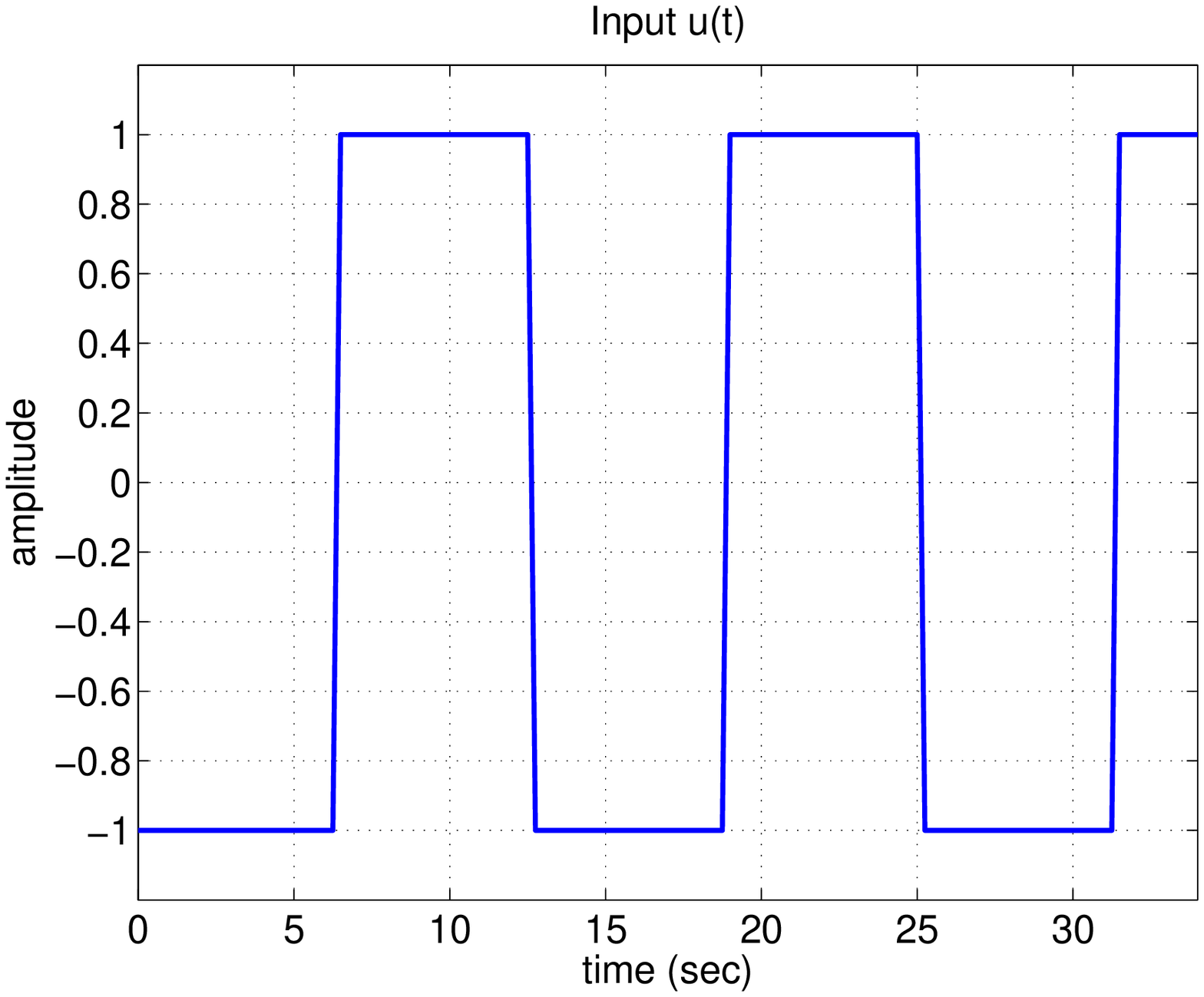}
\caption{Input signal $u(t)$.}
\label{fig:input}
\end{figure}

\begin{figure}[tb]
\centering
\includegraphics[width=\linewidth]{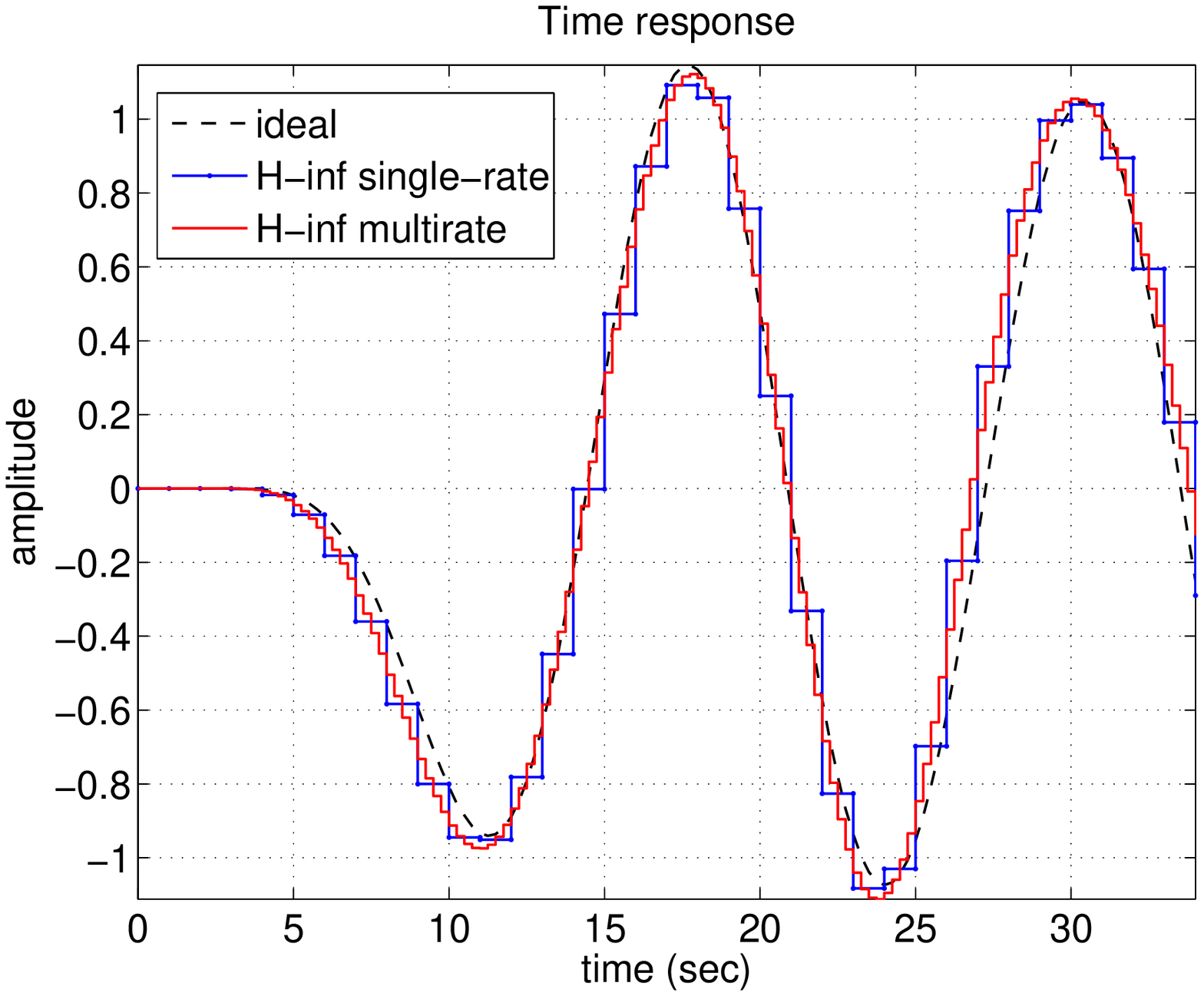}
\caption{Time response: delayed ideal response $y=(Gu)(\cdot - mh)$ (dashed),
response of $H^\infty$-optimal single-rate system (blue),
and that of $H^\infty$-optimal multirate system (red).
}
\label{fig:time_resp}
\end{figure}

\begin{figure}[tb]
\centering
\includegraphics[width=\linewidth]{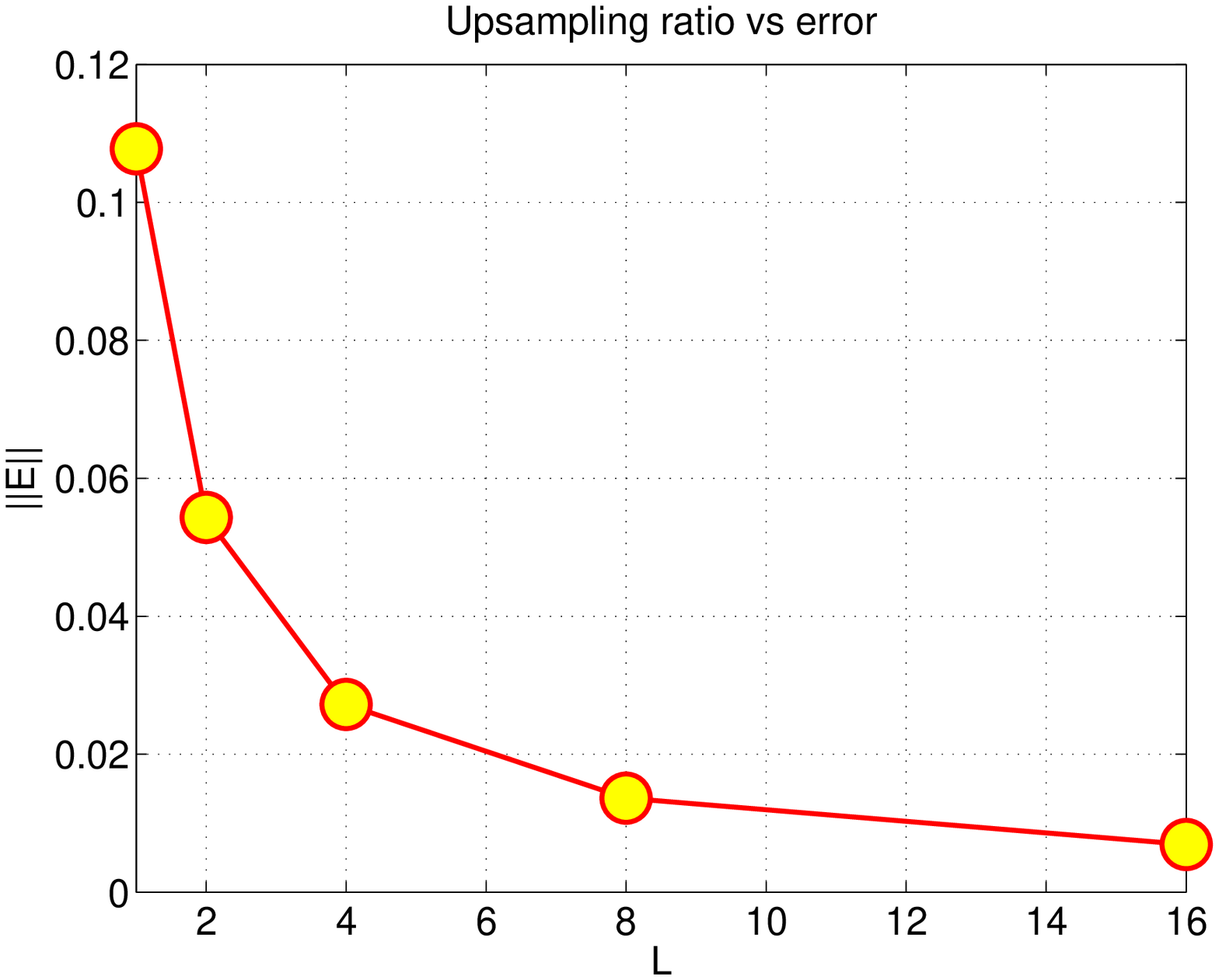}
\caption{Upsampling ratio $L$ versus approximation error $\|\E\|_\infty$.}
\label{fig:LvsPerformance}
\end{figure}

Then we consider an advantage of multirate systems.
We compare the sampled-data $H^\infty$-optimal multirate system with
the single-rate one designed above.
We set the upsampling ratio $L=4$.
We simulate time response of reconstructing a rectangular input signal $u(t)$
shown in Fig.~\ref{fig:input}.
Fig.~\ref{fig:time_resp} shows the time response of the single-rate and multirate systems.
The multirate system approximates the ideal response
more precisely with a slow-rate samples $\{u(0),u(1),\dots\}$.
This is an advantage of the multirate system.
In fact, the approximation error becomes smaller
when we use larger $L$.
To show this,
we take upsampling ratio as $L=1,2,4,8,16$, and compute the $H^\infty$ norm 
of the sampled-data error system $\E$.
Fig.~\ref{fig:LvsPerformance} shows the result.
The figure shows the performance gets better when the upsampling ratio
is increased.

\section{CONCLUSIONS}
\label{sec:conclusions}

In this article, we have proposed a discretization method for analog filters
via sampled-data $H^\infty$ control theory.
The design is formulated as a sampled-data $H^\infty$ optimization problem,
which is approximately reduced to a standard discrete-time $H^\infty$ optimization.
We have also proposed discretization with multirate systems,
which may improve the approximation performance.
We have also discussed feedback controller discretization
with respect to stability.
Design examples show the effectiveness of the proposed method
compared with conventional bilinear transformation and step-invariant transformation.




%
%
\section*{ACKNOWLEDGMENT}
This research is supported in part by the JSPS Grant-in-Aid for Scientific Research (B) No.~24360163
and (C) No.~24560543, and Grant-in-Aid for Exploratory Research No.~22656095.


\end{document}